\documentclass[10pt,twocolumn]{article}     % Enable this line and disable the
\usepackage{amsfonts,amsmath,amssymb,mathrsfs}
\usepackage{color,epsfig,graphicx,graphics}
\usepackage{hyperref}
\usepackage{braket}

\usepackage{times} % assumes new font selection scheme installed

\usepackage{floatrow}
\usepackage{enumerate}

\DeclareMathOperator{\Tr}{Tr}

\DeclareMathOperator{\Sp}{Span}

\begin{document}

\title{An Improved Quantum Projection Filter} % Title, preferably not more

\author{Qing Gao\thanks{School of Automation Science and Electrical Engineering, Beihang University, Beijing 100191, China. (qing.gao.chance@gmail.com)} \and Guofeng Zhang \thanks{Department of Applied Mathematics, Hong Kong Polytechnic University, Hong Kong, China
  (guofeng.zhang@polyu.edu.hk, \url{https://www.polyu.edu.hk/ama/profile/gfzhang/}).}
\and Ian R. Petersen\thanks{Research School of Electrical, Energy and Materials Engineering, The Australian National University, Canberra ACT 2601, Australia.
  (i.r.petersen@gmail.com).}
}
                            % than 10 words.

\maketitle 

%\author[BUAA,BDBC]{Qing Gao}\ead{qing.gao.chance@gmail.com},
%\author[PolyU]{Guofeng Zhang}\ead{Guofeng.Zhang@polyu.edu.hk},
%\author[ANU]{Ian R. Petersen}\ead{i.r.petersen@gmail.com}
%%\author[UDE]{Steven X. Ding}\ead{steven.ding@uni-due.de}    % Add the
%%\author[ADFA]{Ian R. Petersen}\ead{i.r.petersen@gmail.com}               % e-mail address  % (ead) as shown
%
%\address[BUAA]{School of Automation Science and Electrical Engineering, Beihang University, Beijing 100191, China.}
%\address[BDBC]{Beijing Advanced Innovation Center for Big Data and Brain Computing, Beihang University, Beijing 100191, China.} 
%\address[PolyU]{Department of Applied Mathematics, Hong Kong Polytechnic University, Hong Kong SAR, China.} 
%\address[ANU]{Research School of Electrical, Energy and Materials Engineering, The Australian National University, Canberra ACT 2601, Australia.}
%  % Please supply
%

\begin{abstract}                          % Abstract of not more than 200 words.
This work extends the previous quantum projection filtering scheme in [Gao Q., Zhang G., $\&$ Petersen I. R. (2019). An exponential quantum projection filter for open quantum systems. \emph{Automatica}, 99, 59-68.], by adding an optimality analysis result. A reformulation of the quantum projection filter is derived by minimizing the truncated Stratonovich stochastic Taylor expansion of the difference between the true quantum trajectory and its approximation on a lower-dimensional submanifold through quantum information geometric techniques. Simulation results for a qubit example demonstrate better approximation performance for the new quantum projection filter. 
\end{abstract}

\textbf{Keywords.}                         % Five to ten keywords,
Quantum filters; quantum projection filters; Stratonovich stochastic Taylor expansions; quantum information geometry.
                         % keyword list or with the
                                          % help of the Automatica	
                                          % keyword wizard

%%%%%%%%%%%%%%%%%%%%%%%%%%%%%%%%%%%%%%%%%%%%%%%%%%%%%%%%%%%%%%%%%%%%%%%%%%%%%%%%
\section{Introduction}
The quantum filter in the Schr$\ddot{\mbox{o}}$dinger picture, also known as the quantum stochastic master equation, is a stochastic differential equation (SDE) that governs the time evolution of the state of a continuously monitored quantum system (\cite{Belavkin1992}, \cite{Bouten2007}, \cite{Gardiner2000}). In the implementation of many quantum technologies such as measurement-based quantum feedback control \cite{Wiseman2009}, online calculation of the quantum filter equation is an essential requirement which is, however, often computationally demanding when the quantum system under consideration has a high dimensionality (\cite{Gao2016}, \cite{Song2016}). For example, to derive the atomic excitation probability for a two-level atom driven by counter-propagating photons via its quantum filter, a total of 63 scalar valued SDEs has to be numerically solved \cite{Dong2019}. In general, obtaining the solution to the quantum filter equation for an $n-$dimensional quantum system involves solving a system of $n^2-1$ scalar valued SDEs. It is thus particularly useful to develop a computationally more efficient approximation model for the quantum filter equation to speed up the estimation of the conditional quantum state.

In the past decades, research on approximation of the quantum filter equation has been popular, for example, see (\cite{Emzir2016}, \cite{Rouchon2015}, \cite{Tsang2014}). Among the existing results, one interesting approximation scheme is the quantum projection filtering approach, which is motivated by the pioneering work by Brigo, Hanzon and LeGland (\cite{Brigo1998}, \cite{Brigo1999}) where classical (non-quantum) stochastic dynamic systems were considered. The first result on quantum projection filtering can be found in \cite{Handel2005b} which, however, requires exact prior knowledge of an invariant set for the solutions to the quantum filter equation and has limited applications in more complex cases. A less restrictive approach was proposed in \cite{Nielsen2009} where a two-level quantum system was considered. 

In our recent work \cite{Gao2018}, we proposed an exponential quantum projection filtering approach that is applicable to general open quantum systems. The basic idea can be summarized as follows. First, a lower dimensional differential submanifold embedded in the quantum system state space is chosen and endowed with a metric structure. Then, an orthogonal projection operation is defined in terms of this metric and is used to project the coefficients of the quantum filter equation in Stratonovich form to the tangent vector space of the submanifold. In this way, a new curve is generated on the submanifold, which serves as an approximation to the original quantum trajectory. It can be inferred from elementary linear algebra that the coefficients of this new curve are optimal approximations to the coefficients of the quantum filter equation. Nevertheless, it is still questionable how good an approximation the new curve is for the original quantum trajectory.

In this paper, we present an alternative design of quantum projection filters using Stratonovich stochastic Taylor expansions \cite{Kloeden1999} and quantum information geometric techniques \cite{Amari2000}, motivated by the research work in \cite{Armstrong2017} where classical stochastic dynamic systems were considered. This design is optimal in the sense that the truncated Stratonovich stochastic Taylor expansion of the difference between the original quantum trajectory and its approximation on a lower-dimensional submanifold is minimized in the mean square sense. An interesting conclusion is that for a special class of open quantum systems, the proposed method and the one in \cite{Gao2018} yield the same optimal quantum projection filter equation (see Corollary 4.1). We use simulation results from a four-level quantum system example to demonstrate the improved approximation capability of the proposed approach over the one in \cite{Gao2018}, in the general cases.

\textbf{Notation.} The Roman type character i is used to distinguish the imaginary unit i=$\sqrt{-1}$ from the index $i$. For any two $n-$dimensional square matrices $A$ and $B$, $A \otimes B$ means the tensor product of $A$ and $B$ and $[A, B]=AB-BA$ represents the commutator of $A$ and $B$. $A^{\dagger}$ represents the complex conjugate transpose of $A$, $\Tr(A)$ is the trace of matrix $A$, and $\|A\|_F=\sqrt{\Tr(A^{\dagger}A)}$ is the Hilbert-Schmidt norm of matrix $A$. $I_p$ is the $p-$dimensional identity matrix for an integer $p$. $\mathbb{R}^n$ represents the $n$-dimensional real vector space.

\section{The Quantum Filter and The Quantum Projection Filter}
The quantum system model used in this paper follows the one in \cite{Bouten2007}. We consider a finite-dimensional quantum system $\mathcal{Q}$ that has a Hilbert space $\mathbb{H}_{\mathcal{Q}}$ with $\dim \mathbb{H}_{\mathcal{Q}}=n<\infty$, and we use a symmetric Fock space $\mathcal{E}$ to model the quantum bath. The quantum system $\mathcal{Q}$ interacts with the quantum bath on which continuous observations are made through a homodyne detector. The composite system is initially prepared in the state $\rho_{\mbox{total}}=\rho_0\otimes \rho_{\mbox{vacuum}}$, where $\rho_0$ is some state on $\mathbb{H}_{\mathcal{Q}}$ and $\rho_{\mbox{vacuum}}$ is the vacuum state on $\mathcal{E}$. 

By using quantum filtering theory, the dynamics of the quantum density operator of $\mathcal{Q}$ conditioned on the past history of the observation process satisfy the following $It\hat o$ quantum SDE (\cite{Belavkin1992}, \cite{Bouten2007}):
\begin{align}
d\rho_t=&\mathscr{L}_{L, H}^{\dagger}(\rho_t)dt+(L\rho_t+\rho_tL^{\dagger}-\rho_t \Tr(\rho_t(L+L^{\dagger})))\nonumber\\
&\hspace{2.5cm}\times\left(dY_t-\Tr(\rho_t(L+L^{\dagger})dt\right), \label{sprojection1}
\end{align}
where $H$ is the quantum system Hamiltonian, $L$ is the coupling strength operator, $\mathscr{L}_{L, H}^{\dagger}$ is the adjoint of the Lindblad generator $\mathscr{L}_{L, H}(X)=-\mbox{i}[H,X]+LXL^{\dagger}-\frac{1}{2}(L^{\dagger}LX+XL^{\dagger}L)$, and $Y_t$ is the Wiener type classical photocurrent signal generated by the homodyne detector. 

The nonlinear SDE (\ref{sprojection1}) is known as the quantum filter or the quantum stochastic master equation. In the rest of this paper, however, we will work on the following equivalent Stratonovich type unnormalized linear form of (\ref{sprojection1}) for the sake of easy manipulation \cite{Gao2018}:
\begin{eqnarray}
d\bar \rho_t=\left(-\mbox{i}[H, \bar \rho_t]-\mathscr{S}_{L}(\bar \rho_t)\right)dt+\left(L\bar \rho_t+\bar \rho_tL^{\dagger}\right)\circ dY_t, \label{sprojection2}
\end{eqnarray}
where $\mathscr{S}_{L}(\bar \rho_t)=\frac{\left(L+L^{\dagger})L\bar \rho_t+\bar \rho_tL^{\dagger}(L+L^{\dagger}\right)}{2}$, and $\bar \rho_t$ satisfies $\rho_t=\bar \rho_t/\Tr(\bar \rho_t)$ and $\bar \rho_0=\rho_0$. 

In many quantum technologies such as quantum state dependent feedback control, online acquisition of the quantum state $\bar \rho_t$ from (\ref{sprojection2}) is essential. However, this numerical process is often computationally expensive in practice, especially for high-dimensional quantum systems. In fact, a system of $n^2$ scalar valued SDEs has to be solved in order to determine $\bar \rho_t$ for an $n-$dimensional quantum system. In \cite{Gao2018}, a quantum projection filtering approach was proposed, by which an approximation to $\bar \rho_t$ can be obtained by solving a smaller number of SDEs. In the remainder of this section, we will briefly introduce this approach and provide the motivation for our study. 

%\begin{figure}
%\includegraphics[width=1\linewidth]{Projection}
%\caption{Cartoon illustrating the basic setup of the quantum projection filtering strategy. 
%}
%\end{figure}

It can be obtained that the state space of (\ref{sprojection2}) is given by:
\begin{eqnarray}
\mathbb{Q}=\{\bar \rho|\bar \rho\geq 0, \bar \rho=\bar \rho^{\dagger}\},\label{sprojection3}
\end{eqnarray}
which can be naturally regarded as a real differential manifold with dimension $\mbox{dim}(\mathbb{Q})=n^2$. The tangent vector space of $\mathbb{Q}$ is identical with the set
\begin{eqnarray}
\mathbb{A}=\{A|A=A^{\dagger}\}. \label{sprojection4}
\end{eqnarray}
In this way, the solution $\bar{\rho}_t$ to (\ref{sprojection2}) can be geometrically interpreted as a curve on $\mathbb{Q}$ starting from $\rho_0$, and the coefficients of (\ref{sprojection2}), i.e., $-\mbox{i}[H, \bar \rho_t]-\mathscr{S}_{L}(\bar \rho_t)$ and $L\bar \rho_t+\bar \rho_tL^{\dagger}$ are tangent vectors along this curve. Next we choose an $m-$dimensional differential submanifold $\mathbb{S}$ embedded in $\mathbb{Q}$ and suppose that $\mathbb{S}$ can be covered by a single coordinate chart $(\mathbb{S},\theta=(\theta_1,...,\theta_m)^T\in \Theta)$, where $m\leq n^2$ is a positive integer and $\Theta$ is an open subset of $\mathbb{R}^{m}$ containing the origin. The key task of the approximation strategy of quantum projection filtering is to construct a real curve $\theta_t$ on $\Theta$ satisfying the following $m-$dimensional Stratonovich SDE such that the corresponding curve $\bar{\rho}_{\theta_t}$ on $\mathbb{S}$ serves as an approximation to $\bar{\rho}_t$:
\begin{eqnarray}
d\theta_t=f(\theta_t)dt+g(\theta_t)\circ dY_t, \label{sprojection5}
\end{eqnarray}
where $f, g: \mathbb{R}^{m}\rightarrow \mathbb{R}^{m}$ are two real vector valued functions and are tangent vectors along the real curve $\theta_t$.  Equation (\ref{sprojection5}) is then named the \emph{quantum projection filter} in this paper. 

Let $\mathscr{T}_{\bar \rho_{\theta}}(\mathbb{S})$ denote the tangent vector space to $\mathbb{S}$ at each point $\bar \rho_{\theta}$ and let $\{\bar \partial_1,...,\bar \partial_m\}$ be a natural basis of $\mathscr{T}_{\bar \rho_{\theta}}(\mathbb{S})$. That is,
\begin{eqnarray}
\mathscr{T}_{\bar \rho_{\theta}}(\mathbb{S})=\Sp\{\bar \partial_i, i=1,...,m\}. \label{sprojection6}
\end{eqnarray}
In addition, let $\ll,\gg_{\bar \rho_{\theta}}$ be a family of inner products on $\mathbb{A}$ which depends smoothly on $\bar \rho_{\theta}$. Then, we can define a quantum Riemannian metric $r=\left<,\right>_{\bar \rho_{\theta}}$ on $\mathbb{S}$ based on this type of inner product and denote each component of the metric by $r_{ij}(\theta)=\left<\bar \partial_i,\bar \partial_j\right>_{\bar \rho_{\theta}}$. A detailed description of the quantum Riemannian metric can be found in Sections 2.2 and 3.1 of \cite{Gao2018}, which is also summarized in Section 4.1 in this paper (see (\ref{sprojection33})-(\ref{sprojection35})).

%\begin{eqnarray}
%r_{ij}(\theta)=\left<\bar \partial_i,\bar \partial_j\right>_{\bar \rho_{\theta}}.\label{sprojection7}
%\end{eqnarray}
By using the metric structure of $\mathbb{S}$, an orthogonal projection operation $\Pi_{\bar \rho_{\theta}}$ can be defined for every $\theta\in \Theta$ as follows:
\begin{eqnarray}
\Pi_{\bar \rho_{\theta}}: \mathbb{A} &&\longrightarrow \mathscr{T}_{\bar \rho_{\theta}}(\mathbb{S}) \nonumber\\
\nu && \longmapsto \sum_{i=1}^m\sum_{j=1}^m r^{ij}(\theta) \ll\nu,\bar \partial_j \gg_{\bar \rho_{\theta}}\bar \partial_i, \label{sprojection8}
\end{eqnarray}
where the matrix $\left(r^{ij}(\theta)\right)$ is the inverse of $\left(r_{ij}(\theta)\right)$. Based on elementary linear algebra, this orthogonal projection operation provides the best possible way to approximate the coefficients of (\ref{sprojection2}) using vectors in $\mathscr{T}_{\bar \rho_{\theta}}(\mathbb{S})$. Therefore, a natural consideration is that by designing $f$ and $g$ in (\ref{sprojection5}) as
\begin{eqnarray}
f(\theta)&=&\vartheta_*\Pi_{\bar \rho_{\theta}}\left(-\mbox{i}[H,\bar \rho_{\theta}]-\mathscr{S}_{L}(\bar \rho_{\theta})\right) \nonumber\\
g(\theta)&=&\vartheta_*\Pi_{\bar \rho_{\theta}}\left(L\bar \rho_{\theta}+\bar \rho_{\theta}L^{\dagger}\right)\label{sprojection9}
\end{eqnarray}
where $\vartheta_*$ is the pushforward of the diffeomorphism $\vartheta: \bar \rho_{\theta} \rightarrow \theta$ (\cite{Lee2012}), the resulting curve $\bar{\rho}_{\theta_t}$ on $\mathbb{S}$ is then a good approximation to $\bar \rho_t$. However, it is questionable how good this approximation is. 

In this research, we will extend the approximation strategy sketched above such that the difference between $\bar{\rho}_{\theta_t}$ and $\bar \rho_t$ can be minimized in some sense, by presenting a new approach to the design of quantum projection filters. This new approach can be briefly summarized in two steps. First, the stochastic processes $\bar{\rho}_{\theta_t}$ and $\bar \rho_t$ are both expanded as the sum of a truncated term and a remainder term using the Stratonovich stochastic Taylor expansion. The truncated term consists of multiple Stratonovich integrals with constant integrands, while the remainder term consists of multiple Stratonovich integrals with nonconstant integrands and grows at a rate $O(t^k)$ for an integer $k$ in the mean square sense. Second, an orthogonal projection operation defined as in (\ref{sprojection8}) is used to minimize the difference between the truncated terms of $\bar{\rho}_{\theta_t}$ and $\bar \rho_t$, which leads to a new quantum projection filter design.

\section{The Stratonovich Stochastic Taylor Expansion for Quantum Operator Valued Functions}
In this section, we will introduce the Stratonovich stochastic Taylor expansion for quantum operator valued functions that serves as the key tool for deriving our main results. 

The notation used is from  \cite{Kloeden1999}. For a positive integer $l$, a \emph{multi-index} of length $l$ is defined as
\begin{eqnarray}
\alpha=(\alpha_1,\alpha_2,...,\alpha_l), \label{sprojection10}
\end{eqnarray}
where $\alpha_i\in\{0,1\}$ for $i\in\{1,2,...,l\}$. We let $l(\alpha)$ denote the length of $\alpha$ and $n(\alpha)$ denote the number of zeros in $\alpha$ respectively. Let the set of all multi-indices be denoted by $\mathcal{M}$. For any multi-index $\alpha \in \mathcal{M}$ with $l(\alpha)=l\geq 1$, we define $-\alpha$ and $\alpha-$ be the multi-indices obtained by removing the first and the last elements of $\alpha$, respectively. For any two multi-indices $\alpha, \beta\in \mathcal{M}$ with $l(\alpha)=l$ and $l(\beta)=k$, we define the concatenation operation $*$ on $\mathcal{M}$ by 
\begin{eqnarray}
\alpha * \beta=(\alpha_1,\alpha_2,...,\alpha_l, \beta_1,\beta_2,...,\beta_k). \label{sprojection11}
\end{eqnarray}
For an integer $k\geq 0$, we define a subset $\Lambda_k \subset \mathcal{M}$ by
\begin{eqnarray}
\Lambda_k=\{\alpha\in \mathcal{M}: l(\alpha)+n(\alpha)\leq k\},\label{sprojection12}
\end{eqnarray}
and the \emph{remainder set} $\mathcal{R}(\Lambda_k)$ of $\Lambda_k$ by
\begin{eqnarray}
\mathcal{R}(\Lambda_k)=\{\beta \in \mathcal{M}\backslash \Lambda_k: -\beta\in \Lambda_k\}.\label{sprojection13}
\end{eqnarray}
In particular, $\Lambda_0$ represents an empty multi-index. 

Then, we enumerate stochastic integrals with respect to the Wiener process $Y_t$ in (\ref{sprojection2}) and time $t$ using multi-indices. For any time $t$, we define $Y_t^1:=Y_t$ and $Y_t^0:=t$. Let $t_1\leq t_2$ be two time points, the multi-integral for any function $a$ of time associated with a multi-index $\alpha$ is defined by
\begin{eqnarray}
\mbox{I}_{t_1, t_2}^{\alpha}(a)=\left\{
\begin{aligned}
&a(t_2)&  l(\alpha)=0,\\
&\int_{t_1}^{t_2}\mbox{I}_{t_1, s}^{\alpha-}(a)\circ dY_s^{\alpha_l}& \mbox{otherwise.} 
\end{aligned}
\right. \label{sprojection14}
\end{eqnarray}
One can observe that $\mbox{I}_{t_1, t_2}^{\alpha}(a)$ is an $l(\alpha)-$fold integral.

Now, we are ready to formulate the Stratonovich stochastic Taylor expansion for the quantum operator valued functions $\bar \rho_{\theta_{t}}$ and $\bar \rho_t$ introduced in Section 2. Let the derivative of $\bar \rho_{\theta}$ with respect to the real vector $\theta$ be defined by
\begin{eqnarray}
\mathcal{D}_{\theta}\bar \rho_{\theta}=\left(\frac{\partial \bar \rho_{\theta}}{\partial \theta_1}, \frac{\partial \bar \rho_{\theta}}{\partial \theta_2},...,\frac{\partial \bar \rho_{\theta}}{\partial \theta_m}\right),\label{sprojection15}
\end{eqnarray}
and the corresponding $i$th order derivative by 
\begin{eqnarray}
\mathcal{D}_{\theta^i}^i\bar \rho_{\theta}=\underbrace{\mathcal{D}_{\theta}(\mathcal{D}_{\theta}(...(\mathcal{D}_{\theta}\bar \rho_{\theta})...))}_{i \mbox{ successive derivative operations}}. \label{sprojection16}
\end{eqnarray}
Then we define two differentiators on $\bar \rho_{\theta}$ with respect to the SDE (\ref{sprojection5}) as 
\begin{eqnarray}
\left\{
\begin{aligned}
&\underline{L}^0(\bar\rho_{\theta})=\mathcal{D}_{\theta}\bar\rho_{\theta}(f(\theta)\otimes {I}_n),\\
&\underline{L}^1(\bar\rho_{\theta})=\mathcal{D}_{\theta}\bar\rho_{\theta}(g(\theta)\otimes {I}_n),
\end{aligned}
\right. \label{sprojection17}
\end{eqnarray}
respectively. According to the chain rule in Stratonovich stochastic calculus, one has
\begin{eqnarray}
d\bar\rho_{\theta_t}=\underline{L}^0(\bar\rho_{\theta_t})dt+\underline{L}^1(\bar\rho_{\theta_t})\circ dY_t. \label{sprojection18}
\end{eqnarray}
Based on (\ref{sprojection17}), we define a differential operator for $\bar\rho_{\theta}$ associated with a multi-index $\alpha$ as
\begin{eqnarray}
\underline{L}_{\alpha}(\bar\rho_{\theta})=\left\{
\begin{aligned}
&\bar\rho_{\theta}&  l(\alpha)=0,\\
&\underline{L}^{\alpha_1}(\underline{L}_{-\alpha}(\bar\rho_{\theta}))& \mbox{otherwise.} 
\end{aligned}
\right. \label{sprojection19}
\end{eqnarray}
Similarly, considering the SDE (\ref{sprojection2}), we define a differential operator for $\bar \rho$ associated with a multi index $\alpha$ as
\begin{eqnarray}
\underline{D}_{\alpha}(\bar \rho)=\left\{
\begin{aligned}
&\bar \rho &  l(\alpha)=0,\\
&\underline{D}^{\alpha_1}(\underline{D}_{-\alpha}(\bar \rho))& \mbox{otherwise} ,
\end{aligned}
\right. \label{sprojection20}
\end{eqnarray}
with $\underline{D}^0(\bar \rho)=-\mbox{i}[H, \bar \rho]-\mathscr{S}_{L}(\bar \rho)$ and $\underline{D}^1(\bar \rho)=L\bar \rho+\bar \rho L^{\dagger}$. Accordingly, the unnormalized quantum filter equation in (\ref{sprojection2}) can be rewritten as
\begin{eqnarray}
d\bar\rho_t=\underline{D}^0(\bar \rho_t)dt+\underline{D}^1(\bar \rho_t)\circ dY_t. \label{xiugai1}
\end{eqnarray}
Suppose that all the necessary derivatives exist for $\bar \rho_{\theta}$. Here we list the explicit formulations for the differential operators in (\ref{sprojection19}) and (\ref{sprojection20}) with respect to several different multi-indices respectively, which will be useful in later analysis. We have
\begin{eqnarray}
\left\{
\begin{aligned}
&\underline{L}_{(0)}(\bar\rho_{\theta})=\underline{L}^0(\bar\rho_{\theta})=\mathcal{D}_{\theta}\bar\rho_{\theta}(f(\theta)\otimes {I}_n),\\
&\underline{D}_{(0)}(\bar\rho)=\underline{D}^0(\bar\rho)=-\mbox{i}[H, \bar \rho]-\mathscr{S}_{L}(\bar \rho) ,
\end{aligned}
\right. \label{es1}
\end{eqnarray}
\begin{eqnarray}
\left\{
\begin{aligned}
&\underline{L}_{(1)}(\bar\rho_{\theta})=\underline{L}^1(\bar\rho_{\theta})=\mathcal{D}_{\theta}\bar\rho_{\theta}(g(\theta)\otimes {I}_n),\\
&\underline{D}_{(1)}(\bar\rho)=\underline{D}^1(\bar\rho)=L\bar \rho+\bar \rho L^{\dagger},
\end{aligned}
\right. \label{es2}
\end{eqnarray}
and
\begin{eqnarray}
\left\{
\begin{aligned}
\underline{L}_{(1,1)}(\bar\rho_{\theta})=&\underline{L}^1(\underline{L}^1(\bar \rho_{\theta}))\\
=&\mathcal{D}_{\theta}\bar\rho_{\theta}\left(\left(g^T(\theta)\frac{\partial g(\theta)}{\partial \theta}\right)\otimes I_n\right)\\
&+\mathcal{D}_{\theta^2}^2\bar \rho_{\theta}((g(\theta)\otimes g(\theta))\otimes I_n),\\
\underline{D}_{(1,1)}(\bar\rho)=&\underline{D}^1(\underline{D}^1(\bar \rho))\\
=&LL\rho+2L\rho L^{\dagger}+\rho LL^{\dagger}.
\end{aligned}
\right. \label{es3}
\end{eqnarray}
Following Section 3 in \cite{Gao2018}, we let $\mathcal{P}$ denote the measure  by which $Y(t)$ is a Wiener process with zero drift. In the remainder of this paper, all classical random variables are defined in the classical probability space $(\Omega, \mathcal{F}, \mathcal{P})$ while $\mathbb{E}$ represents the expectation operation with respect to $\mathcal{P}$. We introduce the following definition.

\textbf{Definition 3.1.} Let $t_1$ and $t_2$ be two stopping times with 
\begin{eqnarray}
0\leq t_1(\omega)\leq t_2(\omega)\leq T, \label{sprojection21}
\end{eqnarray}
with probability 1, where $\omega\in \Omega$. The order $k$ Stratonovich stochastic Taylor expansions of $\bar \rho_{\theta_{t_2}}$ and $\bar \rho_{t_2}$ at time $t_1$ are given by
\begin{eqnarray}
\left\{
\begin{aligned}
&\mbox{SE}(\bar \rho_{\theta_{t_2}})_{k}=\sum_{\alpha \in \Lambda_k} \mbox{I}_{t_1, t_2}^{\alpha}(1)(\underline{L}_{\alpha}(\bar \rho_{\theta})|_{t_1}),\\
&\mbox{SE}(\bar \rho_{t_2})_k=\sum_{\alpha \in \Lambda_k} \mbox{I}_{t_1, t_2}^{\alpha}(1)(\underline{D}_{\alpha}(\bar \rho)|_{t_1}),
\end{aligned}
\right. \label{sprojection22}
\end{eqnarray}
respectively.

Then we have the following strong convergence result.

\textbf{Theorem 3.1.} Let $k\geq 0$ be an integer. Suppose that there are two constants $R$ and $\bar R$ such that $\mathbb{E}\|\underline{L}_{\beta}(\bar \rho_{\theta_t})\|_F^2\leq R$ and $\mathbb{E}\|\underline{D}_{\beta}(\bar \rho_t)\|_F^2\leq \bar R$ for all time $t$ and any multi-index $\beta \in \mathcal{R}(\Lambda_k)$. Then the following strong convergence capability of the order $k$ Stratonovich stochastic Taylor expansions in (\ref{sprojection22}) holds:
\begin{eqnarray}
\mathbb{E}\left\|\bar \rho_{\theta_{t_2}}-\mbox{SE}(\bar \rho_{\theta_{t_2}})_{k}\right\|_F^2\leq R(2(t_2-t_1))^{k+1}, \label{sprojection23}
\end{eqnarray}
and
\begin{eqnarray}
\mathbb{E}\left\|\bar \rho_{t_2}-\mbox{SE}(\bar \rho_{t_2})_k\right\|_F^2 \leq \bar R(2(t_2-t_1))^{k+1},\label{sprojection24}
\end{eqnarray}
when $t_2-t_1<1$.

\emph{Proof.} See Appendix.

\section{A New Design of Quantum Projection Filters}
In this section, we present a new design of quantum projection filters, by which the difference between $\bar \rho_t$ and $\bar \rho_{\theta_t}$ is minimized in some sense. To be precise, first let us explain how this new approximation scheme is developed using the Stratonovich stochastic Taylor expansion introduced in Section 3. For any time instant $t$, we denote the norm of approximation error by
\begin{eqnarray}
e_t=\mathbb{E}\left\|\bar \rho_t-\bar \rho_{\theta_t}\right\|_F^2,\label{sprojection25}
\end{eqnarray}
which satisfies the following inequality within a small time horizon starting from time $t$ according to Theorem 3.1:
\begin{eqnarray}
&&e_{t+\delta}=\mathbb{E}\left\|\bar \rho_{t+\delta}-\bar\rho_{\theta_{t+\delta}}\right\|_F^2\nonumber\\
&=&\mathbb{E}\left\|\sum_{\alpha \in \Lambda_k} \mbox{I}_{t, t+\delta}^{\alpha}(1)(\underline{L}_{\alpha}(\bar\rho_{\theta_t})-\underline{D}_{\alpha}(\bar \rho_t))+\bar\rho_{\theta_{t+\delta}}-\bar \rho_{t+\delta}\right.\nonumber\\
&&\left.-\sum_{\alpha \in \Lambda_k} \mbox{I}_{t, t+\delta}^{\alpha}(1)(\underline{L}_{\alpha}(\bar \rho_{\theta_t}))+\sum_{\alpha \in \Lambda_k} \mbox{I}_{t, t+\delta}^{\alpha}(1)(\underline{D}_{\alpha}(\bar \rho_t))\right\|_F^2\nonumber\\
&\leq&2\mathbb{E}\left\|\sum_{\alpha \in \Lambda_k\backslash \Lambda_0} \mbox{I}_{t, t+\delta}^{\alpha}(1)(\underline{L}_{\alpha}(\bar \rho_{\theta_t})-\underline{D}_{\alpha}(\bar \rho_t))\right\|_F^2\nonumber\\
&&+2e_t+2(\bar R+R)(2\delta)^{k+1}, \label{sprojection26}
\end{eqnarray}
with $e_0=0$, where $0<\delta<<1$ is a small time perturbation. 

The approximation strategy is implemented as follows. First, the real quantum trajectory $\bar \rho_t$ is evaluated at $\bar \rho_{\theta_t}$. Then based on (\ref{sprojection26}), the two coefficients $f$ and $g$ at time $t$ of the quantum projection filter (\ref{sprojection5}) are determined by solving the following optimization problem for an integer $k\geq 1$:

\emph{Problem 4.1:} \\
$\min\limits_{f, g}
\mathbb{E}\left\|\sum \limits_{\alpha \in \Lambda_k\backslash \Lambda_0} \mbox{I}_{t, t+\delta}^{\alpha}(1)(\underline{L}_{\alpha}(\bar \rho_{\theta_t})-\underline{D}_{\alpha}(\bar \rho_{\theta_t}))\right\|_F^2$.

In this way, the mean square approximation error has a remainder term that is $O(t^{3/2})$.

With the aid of the orthogonal projection operation defined in (\ref{sprojection8}), one obtains the following reformulation of the quantum projection filter by solving Problem 4.1.

\textbf{Theorem 4.1.} The solution to Problem 4.1 with $k=1$ is
\begin{eqnarray}
g(\theta_t)=\vartheta_*\Pi_{\bar \rho_{\theta_t}}\left(L\bar \rho_{\theta_t}+\bar \rho_{\theta_t}L^{\dagger}\right).  \label{sprojection27}
\end{eqnarray}
By choosing the coefficient $g$ as in (\ref{sprojection27}), the following coefficient $f$ solves Problem 4.1 with $k=2$:
\begin{eqnarray}
f(\theta_t)=\vartheta_*\Pi_{\bar \rho_{\theta_t}}\left(\mathscr{L}_{L, H}^{\dagger}(\bar \rho_{\theta_t})-\frac{\underline{L}^1(\underline{L}^1(\bar \rho_{\theta_t}))}{2}\right),\label{sprojection28}
\end{eqnarray}
where $\underline{L}^1(\underline{L}^1(\bar \rho_{\theta}))$ is given in (\ref{es3}).

\emph{Proof.} First, we consider the case that $k=1$. It follows from (\ref{sprojection5}) that the stochastic process $\bar \rho_{\theta_t}$ is independent of the increment $Y_{t+\delta}-Y_t$. Thus we have
\begin{eqnarray}
&&\mathbb{E}\left\|\sum \limits_{\alpha \in \Lambda_1\backslash \Lambda_0} \mbox{I}_{t, t+\delta}^{\alpha}(1)(\underline{L}_{\alpha}(\bar \rho_{\theta_t})-\underline{D}_{\alpha}(\bar \rho_{\theta_t}))\right\|_F^2\nonumber\\
&=&\mathbb{E}\left\|\underline{L}_{(1)}(\bar \rho_{\theta_t})-\underline{D}_{(1)}(\bar \rho_{\theta_t})\right\|_F^2\mathbb{E}(Y_{t+\delta}-Y_t)^2\nonumber\\
&=&\mathbb{E}\left\|\underline{L}^1(\bar \rho_{\theta_t})-\underline{D}^1(\bar \rho_{\theta_t})\right\|_F^2\delta. \label{sprojection29}
\end{eqnarray}
One observes that $\underline{L}^1(\bar \rho_{\theta_t})\in \mathscr{T}_{\bar \rho_{\theta}}(\mathbb{S})$ while $\underline{D}^1(\bar \rho_{\theta_t})\in \mathbb{A}$. Thus, by using the orthogonal projection operation defined in (\ref{sprojection8}), (\ref{sprojection29}) is minimized if the coefficient $g$ is chosen such that
\begin{eqnarray}
\underline{L}^1(\bar \rho_{\theta_t})&=&\mathcal{D}_{\theta_t}\bar\rho_{\theta_t}(g\otimes {I}_n)=(\vartheta^{-1})_*(g)\nonumber\\
&=&\Pi_{\bar \rho_{\theta_t}}\left(\underline{D}^1(\bar \rho_{\theta_t})\right)=\Pi_{\bar \rho_{\theta_t}}\left(L\bar \rho_{\theta_t}+\bar \rho_{\theta_t}L^{\dagger}\right),\label{sprojection30}
\end{eqnarray}
which yields (\ref{sprojection27}).

For the case that $k=2$, one has $\Lambda_2\backslash \Lambda_0=\{(0), (1), (1,1)\}$. It follows from stochastic calculus that $\delta, Y_{t+\delta}-Y_t$ and $\frac{(Y_{t+\delta}-Y_t)^2-\delta}{2}$ are orthogonal with respect to the expectation operation $\mathbb{E}$ \cite{Kloeden1999}. Thus, we have
\begin{eqnarray}
&&\mathbb{E}\left\|\sum \limits_{\alpha \in \Lambda_2\backslash \Lambda_0} \mbox{I}_{t, t+\delta}^{\alpha}(1)(\underline{L}_{\alpha}(\bar \rho_{\theta_t})-\underline{D}_{\alpha}(\bar \rho_{\theta_t}))\right\|_F^2\nonumber\\
&=&\mathbb{E}\Bigg\|\left(\underline{L}^0(\bar \rho_{\theta_t})-\underline{D}^0(\bar \rho_{\theta_t})\right)\delta\nonumber\\
&&+\left(\underline{L}^1(\bar \rho_{\theta_t})-\underline{D}^1(\bar \rho_{\theta_t})\right)(Y_{t+\delta}-Y_t)\nonumber\\
&&\left.+\left(\underline{L}^1(\underline{L}^1(\bar \rho_{\theta_t}))-\underline{D}^1(\underline{D}^1(\bar \rho_{\theta_t}))\right)\frac{(Y_{t+\delta}-Y_t)^2}{2}\right\|_F^2\nonumber\\
&=&\mathbb{E}\left(\left\|\mathscr{M}(\theta_t)\right\|_F^2+\frac{\left\|\mathscr{R}_1(\theta_t)\right\|_F^2}{\delta}+\frac{\left\|\mathscr{R}_2(\theta_t)\right\|_F^2}{4}\right)\delta^2,\label{sprojection31}
\end{eqnarray}
where $\mathscr{M}(\theta_t)=\underline{L}^0(\bar \rho_{\theta_t})-\underline{D}^0(\bar \rho_{\theta_t})+\frac{\underline{L}^1(\underline{L}^1(\bar \rho_{\theta_t}))-\underline{D}^1(\underline{D}^1(\bar \rho_{\theta_t}))}{2}$, $\mathscr{R}_1(\theta_t)=\underline{L}^1(\bar \rho_{\theta_t})-\underline{D}^1(\bar \rho_{\theta_t})$ and $\mathscr{R}_2(\theta_t)=\underline{L}^1(\underline{L}^1(\bar \rho_{\theta_t}))-\underline{D}^1(\underline{D}^1(\bar \rho_{\theta_t}))$. Since $\mathscr{R}_1(\theta_t)$ and $\mathscr{R}_2(\theta_t)$ are both independent of the coefficient $f$, 
$\underline{L}^0(\bar \rho_{\theta_t})\in \mathscr{T}_{\bar \rho_{\theta}}(\mathbb{S})$ and $\underline{D}^0(\bar \rho_{\theta_t})-\frac{\underline{L}^1(\underline{L}^1(\bar \rho_{\theta_t}))-\underline{D}^1(\underline{D}^1(\bar \rho_{\theta_t}))}{2}\in \mathbb{A}$, the solution to Problem 4.1 with $k=2$ is determined by
\begin{eqnarray}
&&\underline{L}^0(\bar \rho_{\theta_t})=\mathcal{D}_{\theta_t}\bar\rho_{\theta_t}(f\otimes {I}_n)=(\vartheta^{-1})_*(f)\nonumber\\
&=&\Pi_{\bar \rho_{\theta_t}}\left(\underline{D}^0(\bar \rho_{\theta_t})-\frac{\underline{L}^1(\underline{L}^1(\bar \rho_{\theta_t}))-\underline{D}^1(\underline{D}^1(\bar \rho_{\theta_t}))}{2}\right)\nonumber\\
&=&\Pi_{\bar \rho_{\theta_t}}\left(\mathscr{L}_{L, H}^{\dagger}(\bar \rho_{\theta_t})-\frac{\underline{L}^1(\underline{L}^1(\bar \rho_{\theta_t}))}{2}\right),\label{sprojection32}
\end{eqnarray}
which yields (\ref{sprojection28}). The proof is completed. $\hspace{2cm}\Box$

\textbf{Remark 4.1.} One observes that the quantum projection filter designed in Theorem 4.1 is \emph{different} from the one in Theorem 3.1 of \cite{Gao2018} (see (\ref{sprojection9}) in this paper). To be specific, the diffusion $g$ in (\ref{sprojection27}) remains the same but the drift $f$ in (\ref{sprojection28}) is different, compared with those in (\ref{sprojection8}). This is because of the optimality criterion that we considered. In Section 5, we will use simulation results to show that this new quantum projection filter has better approximation capabilities than the one formulated in (\ref{sprojection9}).

Here, we present a corollary of Theorem 4.1 which will be useful later. 

\textbf{Corollary 4.1.} The design in Theorem 4.1 reduces to the one in (\ref{sprojection9}), if the first order error term in Problem 4.1 vanishes, that is,
\begin{align*}
\mathbb{E}\left\|\sum \limits_{\alpha \in \Lambda_1\backslash \Lambda_0} \mbox{I}_{t, t+\delta}^{\alpha}(1)(\underline{L}_{\alpha}(\bar \rho_{\theta_t})-\underline{D}_{\alpha}(\bar \rho_{\theta_t}))\right\|_F^2=0.
\end{align*}

Given the geometric structure of $\mathbb{S}$, one can obtain an explicit form for the quantum projection filter using Theorem 4.1. Similar to that in \cite{Gao2018}, the differential submanifold $\mathbb{S}$ is chosen to be:
\begin{eqnarray}
\mathbb{S}=\{\bar \rho_{\theta}\}=\left\{e^{\frac{1}{2}\sum_{i=1}^m\theta_iA_i}\rho_0e^{\frac{1}{2}\sum_{i=1}^m\theta_iA_i}\right\}, \label{sprojection33}
\end{eqnarray}
where the self-adjoint submanifold operators $A_i$ are mutual-communicative and are predesigned. Obviously, any curve $\theta_t$ starting from $\theta_0=0$ corresponds to a curve $\bar \rho_{\theta_t}$ on $\mathbb{S}$ starting from $\rho_0$. A natural basis of the tangent vector space $\mathscr{T}_{\bar \rho_{\theta}}(\mathbb{S})$ is given by
\begin{eqnarray}
\bar \partial_i:=\frac{\partial \bar \rho_{\theta}}{\partial \theta_i}=\frac{1}{2}(A_i \bar \rho_{\theta}+\bar \rho_{\theta}A_i), i=1,2,...,m.\label{sprojection34}
\end{eqnarray}
Next, following a similar procedure to that in Section 7.3 of \cite{Amari2000}, we endow $\mathbb{S}$ with a quantum Fisher metric. To be specific, the \emph{symmetrized inner product} is used to define the inner product $\ll,\gg_{\bar \rho_{\theta}}$ on $\mathbb{A}$ in (\ref{sprojection4}) with respect to the point $\bar \rho_{\theta}\in \mathbb{S}$:
\begin{eqnarray}
\ll A,B\gg_{\bar \rho_{\theta}}=\frac{1}{2}\Tr(\bar \rho_{\theta} AB+\bar \rho_{\theta} BA), \forall A, B\in \mathbb{A}.\label{zhangadd1}
\end{eqnarray}
A useful representation, named the $e-representation$ of a tangent vector $X\in \mathscr{T}_{\bar \rho_{\theta}}(\mathbb{S})$, is defined as the self-adjoint operator $X^{(e)}\in \mathbb{A}$ that satisfies
\begin{eqnarray}
\ll X^{(e)}, A\gg_{\bar \rho_{\theta}}=\Tr\left(XA\right), \forall A\in \mathbb{A}. \label{zhangadd2}
\end{eqnarray}
It follows directly from (\ref{zhangadd2}) that $\bar \partial_i^{(e)}=A_i, i=1,2,...,m.$ 

Then a Riemannian metric $r=\left<,\right>$ can be defined on $\mathbb{S}$ with its components given by
\begin{eqnarray}
r_{ij}(\theta)&=&\left<\bar \partial_i,\bar \partial_j\right>_{\bar \rho_{\theta}}=\ll \bar \partial_i^{(e)}, \bar \partial_j^{(e)}\gg_{\bar \rho_{\theta}}\nonumber\\
&=&\Tr(\bar \partial_iA_j), i,j=1,2,...,m.\label{sprojection35}
\end{eqnarray}
Let the $m\times m$ dimensional quantum Fisher information matrix be denoted by $R(\theta)=(r_{ij}(\theta))$. The following design can be obtained based on Theorem 4.1.

\textbf{Theorem 4.2.} Given the geometric structure of $\mathbb{S}$ in (\ref{sprojection33})-(\ref{sprojection35}), the coefficient $g$ designed from Theorem 4.1 is given by
\begin{eqnarray}
g(\theta_t)=R(\theta_t)^{-1}\left[\begin{array}{c}
	\Tr(\bar \rho_{\theta_t}(A_1L+L^{\dagger}A_1))\\
	 \Tr(\bar \rho_{\theta_t}(A_2L+L^{\dagger}A_2))\\
	 \vdots \\
	 \Tr(\bar \rho_{\theta_t}(A_mL+L^{\dagger}A_m))
	\end{array}
	\right],\label{sprojection36}
\end{eqnarray}
while the coefficient $f$ is given by
\begin{align}
f(\theta_t)&=R(\theta_t)^{-1}\Psi(\theta_t), \label{sprojection37}
\end{align}
where $\Psi(\theta_t)$ is an $m-$dimensional column vector of real functions on $\theta_t$ and the $j$th element is given by
\begin{eqnarray}
\Psi_j(\theta_t)&=&\Tr(\bar \rho_{\theta_t}(\mathscr{L}_{L, H}(A_j))+\frac{\partial g_j(\theta_t)}{\partial \theta_t^T}g(\theta_t)\nonumber\\
&&-\frac{1}{2}g(\theta_t)^T\Delta_j g(\theta_t), \label{sprojection38}
\end{eqnarray}
for $j\in \{1,2,...,m\}$. Here $\Delta_j$ is an $m\times m$ matrix of real function on $\theta_t$ and its entries are given by $\Delta_j(p,q)=\Tr(\bar \rho_{\theta_t}A_pA_qA_j), p,q\in \{1,2,...,m\}$.

\emph{Proof. } To prove Theorem 4.2, let us first recall the basic formula of the pushforward operator with respect to a diffeomorphism. Let $X$ be any vector on $\mathscr{T}_{\bar \rho_{\theta}}(\mathbb{S})$. Then the pushforward of $X$ by the diffeomorphism $\vartheta$, which is denoted by 
$\vartheta_* X$, is a tangent vector along the curve $\theta_t$ and can be explicitly formulated by
\begin{eqnarray}
(\vartheta_*X)_{\theta}=\sum_{j=1}^mX(\vartheta^j(\bar \rho_{\theta}))\left.\frac{\partial}{\partial \theta_j}\right |_{\vartheta(\bar \rho_{\theta})},\label{for1}
\end{eqnarray}
where $\vartheta^j(\bar \rho_{\theta})$ represents the $j$th element of $\vartheta(\bar \rho_{\theta})$ which is well defined because $\vartheta(\bar \rho_{\theta})=\theta\in \mathbb{R}^m$. A simple calculation using (\ref{for1}) yields
\begin{eqnarray}
(\vartheta_*\bar \partial_i)_{\theta}=\sum_{j=1}^m\frac{\partial \theta_j}{\partial \theta_i}\left.\frac{\partial}{\partial \theta_j}\right |_{\vartheta(\bar \rho_{\theta})}=u_i, \label{for2}
\end{eqnarray}
where $u_i$ is the $i$th canonical unit column vector in $\mathbb{R}^m$.

It is known that the pushforward $\vartheta_*$ is a linear operator. Based on (\ref{for2}) and by substituting (\ref{sprojection34}) and (\ref{sprojection35}) into (\ref{sprojection27}) in Theorem 4.1, one has
\begin{eqnarray}
&&g(\theta_t)=\vartheta_*\Pi_{\bar \rho_{\theta_t}}\left(L\bar \rho_{\theta_t}+\bar \rho_{\theta_t}L^{\dagger}\right), \nonumber\\
&=&(\mathcal{D}_{\theta}\vartheta)_{\bar \rho_{\theta_t}}\sum_{i=1}^m\sum_{j=1}^m r^{ij}(\theta_t) \Tr(\bar \rho_{\theta_t}(A_jL+L^{\dagger}A_j))\bar \partial_i\nonumber\\
&=&\sum_{i=1}^m\sum_{j=1}^m r^{ij}(\theta_t) \Tr(\bar \rho_{\theta_t}(A_jL+L^{\dagger}A_j))(\mathcal{D}_{\theta}\vartheta)_{\bar \rho_{\theta_t}}\bar \partial_i\nonumber\\
&=&\sum_{i=1}^m\sum_{j=1}^m r^{ij}(\theta_t) \Tr(\bar \rho_{\theta_t}(A_jL+L^{\dagger}A_j))u_i\nonumber\\
&=&\left[\begin{array}{c}
	\sum_{j=1}^m r^{1j}(\theta) \Tr(\bar \rho_{\theta_t}(A_jL+L^{\dagger}A_j))\\
	 \sum_{j=1}^m r^{2j}(\theta) \Tr(\bar \rho_{\theta_t}(A_jL+L^{\dagger}A_j))\\
	 \vdots \\
	 \sum_{j=1}^m r^{mj}(\theta) \Tr(\bar \rho_{\theta_t}(A_jL+L^{\dagger}A_j))
	\end{array}
	\right]\nonumber\\
&=&R(\theta_t)^{-1}\left[\begin{array}{c}
	\Tr(\bar \rho_{\theta_t}(A_1L+L^{\dagger}A_1))\\
	 \Tr(\bar \rho_{\theta_t}(A_2L+L^{\dagger}A_2))\\
	 \vdots \\
	 \Tr(\bar \rho_{\theta_t}(A_mL+L^{\dagger}A_m))
	\end{array}
	\right].\label{for3}
\end{eqnarray}
Similarly, one can obtain from (\ref{sprojection28}) in Theorem 4.1 that 
\begin{eqnarray}
f(\theta_t)=R(\theta_t)^{-1}\left[\begin{array}{c}
	\Tr\left(\left(\mathscr{L}_{L, H}^{\dagger}(\bar \rho_{\theta_t}-\frac{\Xi(\theta_t)}{2}\right)A_1\right)\\
	\Tr\left(\left(\mathscr{L}_{L, H}^{\dagger}(\bar \rho_{\theta_t}-\frac{\Xi(\theta_t)}{2}\right)A_2\right)\\
	 \vdots \\
	 \Tr\left(\left(\mathscr{L}_{L, H}^{\dagger}(\bar \rho_{\theta_t}-\frac{\Xi(\theta_t)}{2}\right)A_m\right)
	\end{array}
	\right],\label{for3}
\end{eqnarray}
which leads to (\ref{sprojection37}). The proof is thus completed.$\hspace{1.4cm}\Box$

\textbf{Remark 4.2.} The quantum projection filter (\ref{sprojection5}) obtained in Theorem 4.1 is equivalent to the following $It\hat o$ type SDE:
\begin{eqnarray}
d\theta_t=\bar f(\theta_t)dt+g(\theta_t) dY_t, 
\end{eqnarray}
where $g(\theta)$ is given in (\ref{sprojection36}) and the coefficient $f$ is given by
\begin{eqnarray}
\bar f(\theta_t)&=R(\theta_t)^{-1}\Gamma(\theta_t), \label{sprojection39}
\end{eqnarray}
where $\Gamma(\theta_t)$ is an $m-$dimensional column vector of real functions on $\theta_t$ and the $j$th element is given by
\begin{eqnarray}
\Gamma_j(\theta_t)=\Tr(\bar \rho_{\theta_t}(\mathscr{L}_{L, H}(A_j))-\frac{1}{2}g(\theta_t)^T\Delta_j g(\theta_t), \label{sprojection40}
\end{eqnarray}
for $j\in \{1,2,...,m\}$.

Now, we consider a special class of open quantum systems with $L=L^{\dagger}$ for which the design can be further simplified. Then $L$ admits a spectral decomposition as $L=\sum_{i=1}^{\bar n(L)}\bar \lambda_i(L)\bar P_{L_i}$, where $\bar n(L)\leq n$ is the number of nonzero eigenvalues of $L$, the set $\{\bar \lambda_i(L)\}$ contains all of the nonzero real eigenvalues of $L$, and $\{\bar P_{L_i}\}$ is a set of projection operators that satisfies $\bar P_{L_j}\bar P_{L_k}=\delta_{jk}P_{L_k}, j,k=1,2,...,\bar n(L)$. We have the following simplified design based on Theorem 4.2.

\textbf{Corollary 4.2.} By further designing the submanifold $\mathbb{S}$ in (\ref{sprojection33}) as 
\begin{eqnarray}
\begin{cases}
m=\bar n(L),\\
A_i=\bar P_{L_i}, i=1,2,...,m,
\end{cases}\label{sprojection41}
\end{eqnarray}
the coefficient $g$ in (\ref{sprojection36}) becomes
\begin{eqnarray}
g(\theta_t)=2[\bar \lambda_1(L), \bar \lambda_2(L),...,\bar \lambda_m(L)]^T,
\end{eqnarray}
and the coefficient $f$ in (\ref{sprojection37}) becomes
\begin{eqnarray}
f(\theta_t)=R(\theta_t)^{-1}\Phi(\theta_t)-2\Xi,
\end{eqnarray}
where $\Xi=[\bar \lambda_1(L)^2, \bar \lambda_2(L)^2,...,\bar \lambda_m(L)^2]^T$, and $\Phi(\theta_t)$ is an $m-$dimensional column vector of real functions on $\theta_t$ and the $j$th element is given by
\begin{eqnarray}
\Phi_j(\theta_t)&=&\Tr(\mbox{i}\bar \rho_{\theta_t}[H, A_j]). \label{sprojection42}
\end{eqnarray}
Moreover, the objective function in Problem 4.1 with $k=1$ is identically zero for all $t\geq 0$.

\textbf{Remark 4.3.} One observes that for an open quantum system with a self-adjoint coupling operator, the quantum projection filter obtained in Corollary 4.2 is \emph{identical} to the one obtained in Theorem 3.2 in \cite{Gao2018}. This is a direct result of Corollary 4.1. This interesting property also leads to the following two important conclusions of the quantum projection filtering approach in \cite{Gao2018}. 

\begin{itemize}
\item The exponential quantum projection filtering scheme in \cite{Gao2018} fails to solve the optimization problem in Problem 4.1 for $k=2$, because it intuitively treats the diffusion term of (\ref{sprojection2}) as a vector field along the quantum trajectory which is not true in general. In order to more accurately describe the geometric structure of a Stratonovich stochastic differential equation, one needs to refer to other mathematical tools such as the 2-jet theory introduced in \cite{Armstrong2018}; and

\item For an open quantum system with a self-adjoint coupling operator, the design method in Theorem 3.2 in \cite{Gao2018} is also optimal in the sense formulated in this paper. 

\end{itemize}

\textbf{Remark 4.4.} It follows from Theorems 3.2 and 3.3 in \cite{Gao2018} that, for an open quantum systems with a self-adjoint coupling operator, the design of quantum projection filter in Corollary 4.2 significantly reduces the approximation error. This is the key reason why the submanifold and its endowed Riemannian metric structure are designed as in (\ref{sprojection33})-(\ref{sprojection35}).

\section{Comparison Study}
In this section, we use a simple four level quantum system \cite{Carreno2017} example with a non-selfadjoint coupling operator to compare the quantum projection filters obtained from Theorem 4.2 and Theorem 3.1 of \cite{Gao2018}. It has been indicated in Remark 4.3 that for the case that the open quantum system has a self-adjoint coupling operator, the proposed approximation scheme (see Corollary 4.1) reduces to Theorem 3.2 in \cite{Gao2018}, the approximation capability of which has been illustrated via simulation results from a spin system with dispersive coupling example (see Section 4 in \cite{Gao2018}).
 
The four level quantum system used has two close together upper states $\left|2\right>$ and $\left|3\right>$, and two lower states $\left|0\right>$ and $\left|1\right>$. The system Hamiltonian and the coupling strength operator of the qubit system are given by
\begin{eqnarray*}
H=0 \mbox{ and } L=\sum_{j=0}^3l_{j}\left|j\right>\left<j\right|+0.3\left|3\right>\left<0\right|
\end{eqnarray*}
respectively, where $\{l_j\}=\{1, -1, 1, -1\}, j=0,...,3$. Initially, the system is placed at the mixed quantum state $1/8*\left|0\right>\left<0\right|+1/8*\left|1\right>\left<1\right|+3/8*\left|2\right>\left<2\right|+3/8*\left|3\right>\left<3\right|$. In order to obtain the real time quantum state $\rho_t$ from (\ref{sprojection1}), one needs to solve a total of \textbf{15} stochastic differential equations in general. 

%It follows from \cite{James2010} that the quantum system state converges to $\left|0\right>\left<0\right|$ as time goes to infinity. 

The submanifold $\mathbb{S}$ is chosen to be of dimension $4$ and the submanifold operators are given by
\begin{align}
A_1&=\mbox{Diag}\{1,0,0,0\} \nonumber\\
A_2&=\mbox{Diag}\{0,1,0,0\} \nonumber\\
A_3&=\mbox{Diag}\{0,0,1,0\} \nonumber\\
A_4&=\mbox{Diag}\{0,0,0,1\}  \label{sprojection39}
\end{align}
respectively. Then based on Theorem 4.2, one can approximately calculate $\rho_t$ using a quantum projection filter consisting of only \textbf{4} stochastic differential equations.

We use the Monte Carlo discretization approach as in \cite{Higham2001} to solve the quantum stochastic differential equations involved. The simulation parameters used are as follows: the simulation interval $t\in [0,T]$ with $T=5$, the normally distributed variance is $\delta t=T/2^{12}$, and the step size is chosen to be $\Delta t=2\delta t$. In particular, in order to simulate the photocurrent signal $Y_t$, the term $dY_t-\Tr(\rho_t(L+L^{\dagger}))dt$ in (\ref{sprojection1}) is replaced by the instantaneous increment of a Wiener process $dW(t)$, according to the result on page 36 in \cite{Bouten2007}. The quantum filter (\ref{sprojection1}) is calculated first and $Y_t$ can be simulated via the equation $dY(t)=\Tr(\rho_t(L+L^{\dagger}))dt+dW(t)$. 

The approximation capability of the quantum projection filter is demonstrated by considering the Hilbert-Schmidt distance between the real quantum state $\rho_t$ and the normalized state $\rho_{\theta_t}=\frac{\bar \rho_{\theta_t}}{\Tr(\bar \rho_{\theta_t})}$ calculated from the quantum projection filter in (\ref{sprojection5}), i.e., $\sqrt{\Tr(\rho_t-\rho_{\theta_t})^2}$. A number of simulations have been conducted and have illustrated the advantages of the proposed approach. Simulation results from one particular experiment are presented in Fig. 1, from which one can observe that the approach in Theorem 4.2 performs better than the one in Theorem 3.1 of \cite{Gao2018}.

\begin{figure}
\includegraphics[width=1\linewidth]{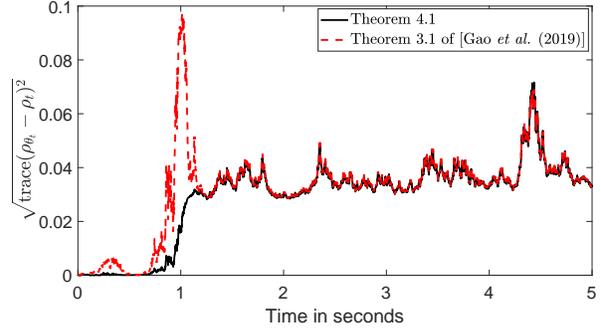}
\caption{Approximation performance comparison.}
\end{figure}

\section{Conclusions}
Based on Stratonovich stochastic Taylor expansions and quantum information geometry techniques, we have provided a new quantum projection filter design. It has been shown by simulation results of a four-level system example that the new quantum projection filter has better approximation capabilities than the existing approach. Future research includes: i) analysis of the approximation errors, especially when the initial quantum state is unavailable; ii) design of more appropriate submanifold for general quantum filter equation and iii) applications of the quantum projection filter to quantum feedback control design.

\section*{Appendix}
\textbf{Proof of Theorem 3.1.} We provide a proof for the convergence capability of the order $k$ Stratonovich stochastic Taylor expansion of $\bar \rho_{\theta_t}$ as shown in (\ref{sprojection23}). The convergence result in (\ref{sprojection24}) can be derived following a similar procedure and the corresponding proof is omitted. 

The proof is decomposed into two steps. 

\emph{Step 1.} We prove
\begin{eqnarray}
\bar \rho_{\theta_{t_2}}=\mbox{SE}(\bar \rho_{\theta_{t_2}})_{k}+\underbrace{\sum_{\alpha \in \mathcal{R}(\Lambda_k)} \mbox{I}_{t_1, {t_2}}^{\alpha}(\underline{L}_{\alpha}(\bar \rho_{\theta})}_{\mbox{Truncation error}}), \label{sproof1}
\end{eqnarray}
by induction on the integer $k$. For the case that $k=0$, the remainder set $\mathcal{R}(\Lambda_k)=\{(0), (1)\}$ and (\ref{sproof1}) becomes 
\begin{align}
\bar \rho_{\theta_{t_2}}=\bar \rho_{\theta_{t_1}}+\int_{t_1}^{t_2}\underline{L}^0(\bar \rho_{\theta_s})ds+\int_{t_1}^{t_2}\underline{L}^1(\bar \rho_{\theta_s})\circ dY_s,\label{sproof2}
\end{align}
which is equivalent to (\ref{sprojection18}).

Let $j\geq 1$ be an integer. It follows from the definitions in (\ref{sprojection12}) and (\ref{sprojection13}) that
\begin{align}
\Lambda_{j+1}\backslash \Lambda_j&=\{\beta\in \mathcal{M}: l(\beta)+n(\beta)=j+1\} \nonumber\\
&=\{\beta\in \mathcal{M}\backslash \Lambda_j: l(-\beta)+n(-\beta)=j, \mbox{ or } j-1\} \nonumber\\
&\subset\{\beta\in \mathcal{M}\backslash \Lambda_j: -\beta \in \Lambda_j\}=\mathcal{R}(\Lambda_j).\label{sproof3}
\end{align}
Now suppose that (\ref{sproof1}) holds for the case that $k=j\geq 1$. By applying the chain rule in Stratonovich stochastic calculus to the function $\underline{L}_{\alpha}(\bar \rho_{\theta})$, one has
\begin{eqnarray}
&&\bar \rho_{\theta_{t_2}}=\mbox{SE}(\bar \rho_{\theta_{t_1}})_{j}+\sum_{\alpha \in \mathcal{R}(\Lambda_j)} \mbox{I}_{t_1, t_2}^{\alpha}(\underline{L}_{\alpha}(\bar \rho_{\theta}))\nonumber\\
&=&\mbox{SE}(\bar \rho_{\theta_{t_1}})_{j}+\sum_{\alpha \in \mathcal{R}(\Lambda_j)\backslash(\Lambda_{j+1}\backslash \Lambda_j)} \mbox{I}_{t_1, t_2}^{\alpha}(\underline{L}_{\alpha}(\bar \rho_{\theta}))\nonumber\\
&&\hspace{0.3cm}+\sum_{\alpha \in \Lambda_{j+1}\backslash \Lambda_j} \mbox{I}_{t_1, t_2}^{\alpha}(\underline{L}_{\alpha}(\bar \rho_{\theta}))\nonumber\\
&=&\mbox{SE}(\bar \rho_{\theta_{t_1}})_{j}+\sum_{\alpha \in \mathcal{R}(\Lambda_j)\backslash(\Lambda_{j+1}\backslash \Lambda_j)} \mbox{I}_{t_1, t_2}^{\alpha}(\underline{L}_{\alpha}(\bar \rho_{\theta}))\nonumber\\
&&\hspace{0.3cm}+\sum_{\alpha \in \Lambda_{j+1}\backslash \Lambda_j} \mbox{I}_{t_1, t_2}^{\alpha}(1)(\underline{L}_{\alpha}(\bar \rho_{\theta})|_{t=t_1}) \nonumber\\
&&\hspace{0.3cm}+\sum_{\alpha \in \Lambda_{j+1}\backslash \Lambda_j}\sum_{z=0}^1\mbox{I}_{t_1, t_2}^{(z)*\alpha}(\underline{L}_{(z)*\alpha}(\bar \rho_{\theta}))\nonumber\\
&=&\mbox{SE}(\bar \rho_{\theta_{t_1}})_{j+1}+\sum_{\alpha \in \bar{\mathcal{R}}} \mbox{I}_{t_1, t_2}^{\alpha}(\underline{L}_{\alpha}(\bar \rho_{\theta})),\label{sproof3}
\end{eqnarray}
where the third equality in (\ref{sproof3}) is obtained by applying the formula (\ref{sproof1}) to the function $\mbox{I}_{t_1, t_2}^{\alpha}(\underline{L}_{\alpha}(\bar \rho_{\theta}))$ for $k=0$, and
\begin{eqnarray}
\bar{\mathcal{R}}&=&[\mathcal{R}(\Lambda_j)\backslash(\Lambda_{j+1}\backslash \Lambda_j)]\bigcup\left[\bigcup_{z=0}^1\{(z)*\alpha : \alpha \in \Lambda_{j+1}\backslash \Lambda_j\}\right]\nonumber\\
&=&\left[\{\beta \in \mathcal{M}\backslash \Lambda_j: -\beta\in \Lambda_j\}\backslash \{\beta \in \mathcal{M}\backslash \Lambda_j: \beta\in \Lambda_{j+1}\}\}\right]\nonumber\\
&&\hspace{1cm}\bigcup \{\beta \in \mathcal{M}: -\beta \in \Lambda_{j+1}\backslash \Lambda_j\}\nonumber\\
&=&\{\beta \in \mathcal{M}\backslash \Lambda_{j+1}: -\beta\in \Lambda_j\}\nonumber\\
&&\hspace{1cm}\bigcup \{\beta \in \mathcal{M}\backslash \Lambda_{j+1}: -\beta \in \Lambda_{j+1}\backslash \Lambda_j\}\nonumber\\
&=&\{\beta \in \mathcal{M}\backslash \Lambda_{j+1}: -\beta\in \Lambda_{j+1}\}=\mathcal{R}(\Lambda_{j+1}).\label{sproof4}
\end{eqnarray}
By combining (\ref{sproof3}) and (\ref{sproof4}), one obtains that (\ref{sproof1}) holds for the case $k=j+1$. Then by mathematical induction, (\ref{sproof1}) holds for any nonnegative integer $k$.

\emph{Step 2.} We prove that the truncation error term in (\ref{sproof1}) satisfies 
\begin{eqnarray}
\mathbb{E}\left\|\sum_{\alpha \in \mathcal{R}(\Lambda_k)} \mbox{I}_{t_1, t_2}^{\alpha}(\underline{L}_{\alpha}(\bar \rho_{\theta}))\right\|_F^2\leq R(2(t_2-t_1))^{k+1}. \label{sproof5}
\end{eqnarray}
It is noted that $l(\alpha)+n(\alpha)\in \{k+1, k+2\}, \forall \alpha \in \mathcal{R}(\Lambda_k)$. It is also noted that the number of multi-indices belonging to $\mathcal{R}(\Lambda_k)$ is $2^{k+1}$. Therefore, it is sufficient to prove (\ref{sproof5}) if we can show that for any $\beta \in \mathcal{R}(\Lambda_k)$
\begin{eqnarray}
\max_{\alpha \in \mathcal{R}(\Lambda_k)} \mathbb{E}\left(\mbox{I}_{t_1, t_2}^{\alpha}(\underline{L}_{\beta}(\bar \rho_{\theta}))\right)\leq R(t_2-t_1)^{l(\alpha)+n(\alpha)}. \label{sproof6}
\end{eqnarray}
Now we prove (\ref{sproof6}) by induction on $l(\alpha)\geq 1$. For the time interval $[t_1,t_2]$, we define a partition of time by $t_1=t^0<t^1<t^2...<t^p=t_2$ where the positive integer $p$ is chosen to be big enough. 

First, consider the case that $l(\alpha)=1$ with $\alpha=(0)$. It then follows from the $H\ddot{o}lder$ inequality that
\begin{eqnarray}
&&\mathbb{E}\left\|\mbox{I}_{t_1, t_2}^{(0)}(\underline{L}_{\beta}(\bar \rho_{\theta}))\right\|_F^2=\left\|\int_{t_1}^{t_2}\underline{L}_{\beta}(\bar \rho_{\theta_s})ds\right\|_F^2\nonumber\\
&=&\left\|\lim\limits_{p\to \infty} \sum_{i=0}^p\underline{L}_{\beta}\left(\bar \rho_{\frac{t^{i}+t^{i+1}}{2}}\right)\left(t^{i+1}-t^i\right)\right\|_F^2\nonumber\\
&\leq&\left\{\lim\limits_{p\to \infty}\sum_{i=0}^p\left\|\underline{L}_{\beta}\left(\bar \rho_{\theta_{\frac{t^{i}+t^{i+1}}{2}}}\right)\right\|_F(t^{i+1}-t^i)\right\}^2\nonumber\\
&=&\left\{\int_{t_1}^{t_2}\|\underline{L}_{\beta}(\bar \rho_{\theta_s})\|_Fds\right\}^2\nonumber\\
&\leq&(t_2-t_1)\left\{\int_{t_1}^{t_2}\|\underline{L}_{\beta}(\bar \rho_{\theta_s})\|_F^2ds\right\}\nonumber\\
&\leq&R(t_2-t_1)^2=R(t_2-t_1)^{l(\alpha)+n(\alpha)}.\label{sproof7}
\end{eqnarray}
On the other hand, considering the case that $l(\alpha)=1$ with $\alpha=(1)$, one has
\begin{eqnarray}
&&\mathbb{E}\|\mbox{I}_{t_1, t_2}^{(1)}(\underline{L}_{(1)}(\bar \rho_{\theta}))\|_F^2=\mathbb{E}\left\|\int_{t_1}^{t_2}\underline{L}_{(1)}(\bar \rho_{\theta_s})\circ dY_s\right\|_F^2\nonumber\\
&=&\mathbb{E}\left\|\lim\limits_{p\to \infty} \sum_{i=0}^p\underline{L}_{(1)}\left(\bar \rho_{\theta_{\frac{t^{i}+t^{i+1}}{2}}}\right)(Y_{t^{i+1}}-Y_{t^i})\right\|_F^2\nonumber\\
&\leq&\lim\limits_{p\to \infty}\sum_{i=0}^p\mathbb{E}\left\{\left\|\underline{L}_{(1)}\left(\bar \rho_{\theta_{\frac{t^{i}+t^{i+1}}{2}}}\right)\right\|_F^2\right\}(t^{i+1}-t^i)\nonumber
\end{eqnarray}
\begin{eqnarray}
&=&\mathbb{E}\left\{\int_{t_1}^{t_2}\|\underline{L}_{(1)}(\bar \rho_{\theta_s})\|_F^2ds\right\}\nonumber\\
&\leq& R(t_2-t_1)=R(t_2-t_1)^{l(\alpha)+n(\alpha)}. \label{sproof8}
\end{eqnarray}
Suppose that (\ref{sproof6}) holds for the case that $l(\alpha)=j\geq 1$. Now let $l(\alpha)=j+1$ with $\alpha_{j+1}=0$. Then $l(\alpha)+n(\alpha)\geq 2n(\alpha)\geq 2$, and $l(\alpha-)+n(\alpha-)=l(\alpha)+n(\alpha)-2$. By using the inductive hypothesis, one has
\begin{eqnarray}
&&\mathbb{E}\|\mbox{I}_{t_1, t_2}^{\alpha}(\underline{L}_{\beta}(\bar \rho_{\theta}))\|_F^2=\mathbb{E}\left\|\int_{t_1}^{t_2}\mbox{I}_{t_1, s}^{\alpha-}(\underline{L}_{\beta}(\bar \rho_{\theta_s}))ds\right\|_F^2\nonumber\\
&\leq&(t_2-t_1)\int_{t_1}^{t_2}\mathbb{E}\left\|\mbox{I}_{t_1, s}^{\alpha-}(\underline{L}_{\beta}(\bar \rho_{\theta_s}))\right\|_F^2ds\nonumber\\
&\leq&(t_2-t_1)\int_{t_1}^{t_2}R(s-t_1)^{l(\alpha)+n(\alpha)-2}ds\nonumber\\
&\leq&\frac{R(t_2-t_1)^{l(\alpha)+n(\alpha)}}{l(\alpha)+n(\alpha)-1}\leq R(t_2-t_1)^{l(\alpha)+n(\alpha)}. \label{sproof9}
\end{eqnarray}
On the other hand, let $\alpha_{j+1}=1$. Then $l(\alpha)+n(\alpha)\geq l(\alpha)\geq 1$, and $l(\alpha-)+n(\alpha-)=l(\alpha)+n(\alpha)-1$. By using the inductive assumption, one has
\begin{eqnarray}
&&\mathbb{E}\|\mbox{I}_{t_1, t_2}^{\alpha}(\underline{L}_{\beta}(\bar \rho_{\theta_t}))\|_F^2\nonumber\\
&=&\mathbb{E}\left\|\int_{t_1}^{t_2}\mbox{I}_{t_1, s}^{\alpha-}(\underline{L}_{\beta}(\bar \rho_{\theta_s}))\circ dY_s\right\|_F^2\nonumber\\
&\leq&\mathbb{E}\left\{\int_{t_1}^{t_2}\|\mbox{I}_{t_1, s}^{\alpha-}(\underline{L}_{\beta}(\bar \rho_{\theta_s}))\|_F^2ds\right\}\nonumber\\
&\leq&\int_{t_1}^{t_2}R(s-t_1)^{l(\alpha)+n(\alpha)-1}ds\nonumber\\
&\leq&\frac{R(t_2-t_1)^{l(\alpha)+n(\alpha)}}{l(\alpha)+n(\alpha)}\leq R(t_2-t_1)^{l(\alpha)+n(\alpha)}.\label{sproof10}
\end{eqnarray}
By combining (\ref{sproof9}) and (\ref{sproof10}), one obtains that (\ref{sproof6}) holds for $l(\alpha)=j+1$. One can then conclude that (\ref{sproof6}) holds for any nonnegative integer $k$ by mathematical induction. $\hspace{1cm}\Box$


\begin{thebibliography}{99}
\bibitem[Amari \& Nagaoka (2000)]{Amari2000}
Amari S. \& Nagaoka H. (2000).
\emph{Methods of Information Geometry.}
Oxford: Oxford University Press.

\bibitem[Armstrong \& Brigo (2019)]{Armstrong2017}
Armstrong J. \& Brigo D. (2019).
Optimal approximation of SDEs on submanifolds: the $It\hat o$-vector and $It\hat o$-jet projection.
\emph{Proceedings of London Mathemtical Society,} 119, 176-213.

\bibitem[Armstrong \& Brigo (2018)]{Armstrong2018}
Armstrong J. \& Brigo D. (2018).
Intrinsic stochastic differential equations as jets.
\emph{Proceedings of the Royal Society of London A: Mathematical, Physical and Engineering Sciences,} 474, http://doi.org/10.1098/rspa.2017.0559.

\bibitem[Belavkin (1992)]{Belavkin1992}
Belavkin V.P. (1992).
Quantum stochastic calculus and quantum nonlinear filtering.
\emph{Journal of Multivariate Analysis}, 42, 171-201. 


\bibitem[Bouten \emph{et al.} (2007)]{Bouten2007}
Bouten L., van Handel R. \& James M. R. (2007).
An introduction to quantum filtering.
\emph{SIAM Journal on Control and Optimization}, 46, 2199-2241.


\bibitem[Brigo \emph{et al.} (1998)]{Brigo1998}
Brigo D., Hanzon B. \& LeGland F. (1998).
A differential geometric approach to nonlinear filtering: the projection filter.
\emph{IEEE Transactions on Automatic Control,} 43, 247-252.


\bibitem[Brigo \emph{et al.} (1999)]{Brigo1999}
Brigo D., Hanzon B. \& LeGland F. (1999).
Approximate filtering by projection on the manifold of exponential densities.
\emph{Bernoulli,} 5, 495-543.

\bibitem[Carre$\tilde{n}$o \emph{et al.} (2017)]{Carreno2017}
Carre$\tilde{n}$o F.,  Ant$\acute{o}$n M. A., Yannopapas V. \& Paspalakis E. (2017).
Control of the absorption of a four-level quantum system near a plasmonic nanostructure.
\emph{Physical Review B}, 95, 195410.

\bibitem[Dong \emph{et al.} (2019)]{Dong2019}
Dong Z., Zhang G. \& Amini N. (2019).
Quantum filtering for a two-level atom driven by two
counter-propagating photons.
\emph{Quantum Information Processing,} 18, 136.

\bibitem[Emzir \emph{et al.} (2017)]{Emzir2016}
Emzir M.F. , Woolley M.J. \& Petersen I.R. (2017).
A quantum extended Kalman filter.
\emph{Journal of Physics A: Mathematical and Theoretical,} 50, 225301.

\bibitem[Gardiner \& Zoller (2000)]{Gardiner2000}
Gardiner C. W. \& Zoller P. (2000).
\emph{Quantum Noise: A Handbook of Markovian and Non-Markovian Quantum Stochastic Methods with Applications to Quantum Optics. 2nd Edition.}
New York: Springer-Verlag. 

\bibitem[Gao \emph{et al.} (2016)]{Gao2016}
Gao Q., Dong D. \& Petersen I.R. (2016).
Fault tolerant quantum filtering and fault detection for quantum systems.
\emph{Automatica,} 71, 125-134.

\bibitem[Gao \emph{et al.} (2019)]{Gao2018}
Gao Q., Zhang G. \& Petersen I.R. (2019).
An exponential quantum projection filter for open quantum systems.
\emph{Automatica,} 99, 59-68.


\bibitem[Higham (2001)]{Higham2001}
Higham D. (2001).
An algorithmic introduction to numerical simulation of stochastic differential equations.
\emph{SIAM Review}, 43, 525-546.


\bibitem[James \& Gough (2010)]{James2010}
James M. R.  \& Gough J. E. (2010).
Quantum dissipative systems and feedback
control design by interconnection.
\emph{IEEE Transactions on Automatic Control,} 55, 1806-1821.


\bibitem[Kloeden \& Platen (1999)]{Kloeden1999}
Kloeden P. E., \& Platen E. (1999).
\emph{Numerical Solution of Stochastic Differential Equations. }
Berlin, New York, Springer.


\bibitem[Lee (2012)]{Lee2012}
Lee, J. M. (2012).
\emph{Introduction to Smooth Manifolds, 2nd ed}.
New York, Springer.


\bibitem[Nielsen \emph{et al.} (2009)]{Nielsen2009}
Nielsen A., Hopkins A. \& Mabuchi H. (2009).
Quantum filter reduction for measurement-feedback control via unsupervised manifold learning.
\emph{New Journal of Physics}, 11, 105043.


\bibitem[Rouchon \& Ralph (2015)]{Rouchon2015}
Rouchon P. \& Ralph J.F. (2015).
Efficient quantum filtering for quantum feedback control.
\emph{Physical Review A}, 91, 012118.


\bibitem[Song \emph{et al.} (2016)]{Song2016}
Song H., Zhang G. \& Xi Z. (2016).
Continuous-mode multi-photon filtering.
\emph{SIAM Journal on Control and Optimization}, 54, 1602-1632.


\bibitem[Tsang (2014)]{Tsang2014}
Tsang M. (2014).
Volterra filters for quantum estimation and detection.
\emph{Physical Review A}, 92, 062119.



\bibitem[van Handel \& Mabuchi (2005)]{Handel2005b}
van Handel R. \& Mabuchi H. (2005).
Quantum projection filter for a highly nonlinear model in cavity QED. 
\emph{Journal of Optics B: Quantum and Semiclassical Optics,} 7, S226-S236.

\bibitem[Wiseman \& Milburn (2010)]{Wiseman2009}
Wiseman H.M. \& Milburn G.J. (2010).
\emph{Quantum Measurement and Control.}
Cambridge, U.K.: Cambridge University Press.





\end{thebibliography}
\end{document}